\documentclass[aps,prl,showpacs,twocolumn,superscriptaddress]{revtex4-1}

\usepackage{bm,color,amsmath,amssymb,latexsym,graphicx,psfrag,accents,float}
\usepackage{amscd,dsfont,wasysym,mathrsfs,mathtools}
\usepackage{multirow}
\usepackage{dcolumn}
\usepackage{xcolor}
\usepackage{comment}

\newcolumntype{L}[1]{>{\raggedright\arraybackslash}p{#1}}
\newcolumntype{C}[1]{>{\centering\arraybackslash}p{#1}}
\newcolumntype{R}[1]{>{\raggedleft\arraybackslash}p{#1}}

			\newcommand{\e}[1]{\begin{align}{#1}\end{align}}

		\newcommand{\f}[2]{\frac{#1}{#2}}

		\newcommand{\la}[1]{\label{#1}}

		\newcommand{\q}[1]{Eq.\ (\ref{#1})}

		\newcommand{\fig}[1]{Fig.\ \ref{#1}}

		\newcommand{\ocite}[1]{Ref.\ \onlinecite{#1}}

		\newcommand{\eq}{=&\;}

		\newcommand{\R}{\mathbb{R}}
		\newcommand{\Z}{\mathbb{Z}}

\newcommand{\nabk}{\nabla_{\boldsymbol{k}}}

\newcommand{\var}{\varepsilon}

\newcommand{\bk}{\boldsymbol{k}}

\newcommand{\bn}{\boldsymbol{n}}

\newcommand{\bA}{\boldsymbol{A}}
\newcommand{\bB}{\boldsymbol{B}}

\newcommand{\bF}{\boldsymbol{F}}

\newcommand{\bze}{\boldsymbol{0}}

\newcommand{\bcalf}{\boldsymbol{\calf}}

\newcommand{\bsigma}{\boldsymbol{\sigma}}

\newcommand{\W}{{\cal W}}

\newcommand{\sx}{\sigma_{\sma{1}}}
\newcommand{\sy}{\sigma_{\sma{2}}}
\newcommand{\sz}{\sigma_{\sma{3}}}

\newcommand{\calc}{{\cal C}}

\newcommand{\calf}{{\cal F}}

\newcommand{\minus}{\text{-}}

\newcommand{\braket}[2]{\big\langle #1 \big| #2 \big\rangle}

\newcommand{\pdg}[1]{{#1}^{\phantom{\dagger}}}

\newcommand{\lin}{\notag \\}

\newcommand{\bpm}{\begin{pmatrix}}
\newcommand{\epm}{\end{pmatrix}}

\newcommand{\bal}{\begin{align}}

\newcommand{\dg}[1]{#1^{\scriptstyle{\dagger}}}

\newcommand{\sma}[1]{\scriptscriptstyle{#1}}

\begin{document}

 \title{Teleportation of Berry curvature on the surface of a Hopf insulator}

   
      \author{A. Alexandradinata}%
         \email{aalexan7@illinois.edu}
   \affiliation{Department of Physics, University of Illinois at Urbana-Champaign, Urbana, Illinois 61801-2918, USA}%
   \author{Aleksandra Nelson}
   \affiliation{Physik-Institut, Universit\"{a}t Z\"{u}rich, Winterthurerstrasse 190 
   CH-8057, Z\"{u}rich, Switzerland}
     \affiliation{Department of Physics, 
    St. Petersburg State University, St. Petersburg, 199034, Russia}
   \author{Alexey A. Soluyanov}
   \affiliation{Physik-Institut, Universit\"{a}t Z\"{u}rich, Winterthurerstrasse 190 
   CH-8057, Z\"{u}rich, Switzerland}
    \affiliation{Department of Physics, 
    St. Petersburg State University, St. Petersburg, 199034, Russia}

\begin{abstract}
The existing paradigm for topological insulators asserts that an energy gap separates conduction and valence bands with opposite topological invariants. Here, we propose that \textit{equal}-energy bands with opposite Chern invariants can be \textit{spatially} separated -- onto opposite facets of a finite crystalline Hopf insulator. On a single facet, the number of curvature quanta is in one-to-one correspondence with the bulk homotopy invariant of the Hopf insulator -- this originates from a novel bulk-to-boundary flow of Berry curvature which is \textit{not} a type of Callan-Harvey anomaly inflow. In the continuum perspective, such nontrivial surface states arise as \textit{non}-chiral, Schr\"odinger-type modes on the domain wall of a generalized Weyl equation -- describing a pair of opposite-chirality Weyl fermions acting as a \textit{dipolar} source of Berry curvature. A rotation-invariant lattice regularization of the generalized Weyl equation manifests a generalized Thouless pump -- which translates charge by one lattice period over \textit{half} an adiabatic cycle, but reverses the charge flow over the next half.
\end{abstract}
\date{\today}

\maketitle

\begin{figure}
    \centering
    \includegraphics[width=\columnwidth]{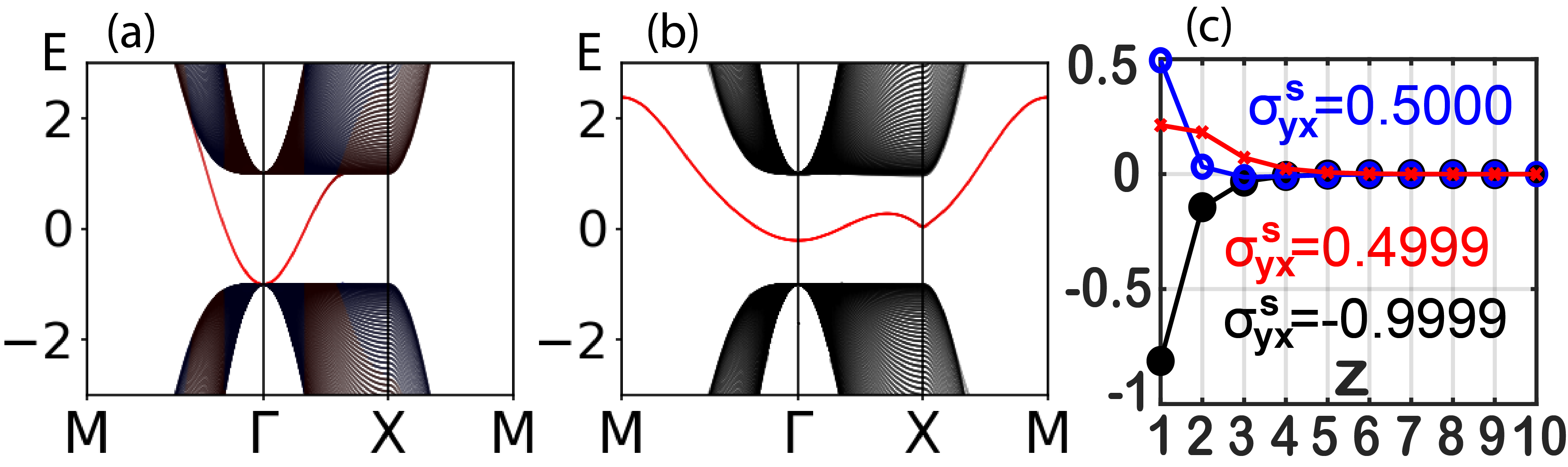}
    \caption{(a) Red line depicts the 001-surface states of the Hopf insulator in the Moore-Ran-Wen tight-binding model\cite{Hopfinsulator_Moore} with $\chi{=}1$ and open boundary conditions. (b) Surface band of (a) is detached by a symmetric deformation of the surface. (c) For a fifty-cell-wide 001 slab, we plot  the layer-resolved Chern density [Eq.\ (12) in \ocite{essin_magnetoelectric}] of the bulk valence band (blue), bulk conduction band (red), and surface band (black) illustrated in (b). Summing the Chern density over ten cells approximates $\sigma^s_{yx}$ (in units of $e^2/h$) to exponential accuracy.} \label{fig:edge}
\end{figure}

The Hopf insulator\cite{Hopfinsulator_Moore} is the paradigm for a topological insulator that is neither a stable topological insulator in the tenfold way,\cite{schnyder_classify3DTIandTSC,kitaev_periodictable} nor a fragile topological insulator\cite{po_fragile,crystalsplit_AAJHWCLL,bouhon_wilsonloopapproach} in topological quantum chemistry.\cite{bradlyn_disconnectedEBR,zhida_fragileaffinemonoid} Having a $\bk$-dependent matrix Hamiltonian that is only a sum of Pauli matrices, the Hopf insulator is conceptually the simplest topological insulator. As originally touted by Moore, Ran and Wen,\cite{Hopfinsulator_Moore} the Hopf insulator is the first-known 3D magnetic bulk-insulator with topological surface states, as illustrated in \fig{fig:edge}(a).  Contrary to some expectations,\cite{protectedHopf_chunxiao,hopfsurfacestates_deng,kennedy_hopfchern} we will demonstrate that all surfaces can be made insulating while preserving the bulk gap [cf.\ \fig{fig:edge}(b)]. What then is `topological' about such surface states which can be spectrally disconnected from the bulk states? 

We will show that  the \textit{surface Chern number} -- defined as the first Chern number of \textit{all} surface-localized bands -- is in one-to-one correspondence with the  integer-valued, bulk homotopy  invariant $\chi$. $\chi$  classifies  maps from the three-dimensional Brillouin zone (BZ) to the Bloch sphere of pseudospin-half wave functions,\cite{pontrjagin_classification,kennedy_hopfchern}  and is given by a gauge-invariant BZ-integral of the Abelian Chern-Simons three-form\cite{wilczekzee_linkingnumbers,Hopfinsulator_Moore}
\e{\chi=-\f1{4\pi^2}\int_{BZ} \bF \cdot \bA \,d^3k,\la{definechi}}
with $\bF{=}\nabla{\times}\bA$, and $\bA(\bk){=}\braket{u}{i\nabk u}$ the Berry connection of the energy-nondegenerate band.\cite{berry_quantalphase}

In the geometric theory of polarization,\cite{zak_berryphase,kingsmith_polarization}  the charge polarization is proportional to $\int_{BZ}\bA$; analogously, one may view $\int_{BZ} \bA\cdot \bF$ as proportional to a real-space polarization of Berry curvature.\cite{essin_magnetoelectric,maryam_adiabaticpumpCSA,olsen_surfacetheoremCSA} Just as the charge polarization  is revealed by the charge that accumulates at a surface termination,\cite{vanderbilt1993} it is known from the theory of magnetoelectric polarizability\cite{qi_topologicalfieldtheory,essin_magnetoelectric,essin_magnetoelectric_erratum,essin_orbitalmagnetoelectric_bandinsulators,malashevich_magnetoelectric,vanderbilt_book_berryelectronicstructure} that a Berry-curvature polarization of $\chi$ contributes $({-}\chi/2) e^2/h$ to the geometric component ($\sigma_{yx}^s$)  of the surface anomalous Hall conductance.\cite{essin_orbitalmagnetoelectric_bandinsulators,rauch_sahc,malashevich_magnetoelectric}  Crucially, the bulk conduction and valence bands individually and equally contribute $({-}\chi/2) e^2/h$, owing to the fundamental antisymmetry of all Pauli-matrix Hamiltonians: $\sy H(\bk)\sy{=}{-}\overline{H(\bk)}$. If all bands in the Hilbert space are accounted for, the net contribution to $\sigma^s_{yx}$ should vanish,\cite{rauch_sahc} hence there should exist surface band(s) which contribute  $\chi e^2/h$ to $\sigma^s_{yx}$. This heuristic deduction is confirmed numerically in  \fig{fig:edge}(c) and will be demystified in this Letter.

It will be shown that such nontrivial surface bands  originate from a bulk-to-boundary flow of Berry curvature --  at the critical point between the trivial and Hopf insulator. Such flow does \textit{not} occur via the Callan-Harvey anomaly inflow mechanism,\cite{callanharvey_anomalies,golterman_chiralfermions} which is understood through chiral zero modes of the Dirac equation with a mass coupling to a domain wall.\cite{jackiwrebbi_solitons,creutz_chiralsymmetry} Instead, the  surface states of the Hopf insulator arise as \textit{non}-chiral, Schr\"odinger-type  modes on the domain wall of a generalized Weyl equation -- describing a pair of opposite-chirality Weyl fermions acting as a dipolar source of Berry curvature. A single surface facet of a Hopf-insulating crystal is \textit{un}describable by a 2D lattice model of a Chern insulator, because the conjugate surface states (with opposite Chern number) lie on the \textit{opposite} surface of a finite slab --  realizing a nonlocal `teleportation' of Berry curvature.



We propose the surface Chern number $C_s$ as a topological invariant for any three-spatial-dimensional crystalline Hamiltonian that is a two-by-two matrix at each $\bk{\in}BZ$, with a spectral gap (at each $\bk$) that separates a low-energy and high-energy bulk band. Both bulk bands are assumed to have trivial first Chern class, meaning  the first Chern number vanishes on any 2D cut of the BZ.\cite{Panati_trivialityblochbundle}  For such class of Hamiltonians which include the Hopf insulator, we consider a surface termination whose reduced Brillouin zone ($rBZ$) is a 2D cut of BZ. Throughout this letter, `bulk' objects or properties refer to the crystalline interior which locally has the symmetry of a 3D space group, up to exponentially-weak corrections due to a surface termination. 

The triviality of the first Chern class implies any surface-localized mode is continuously deformable to a flat-band dispersion\cite{fidkowski_bulkboundary,nogo_AAJH} and is therefore  \textit{detachable} from   bulk  bands. To `detach' means to energetically separate a surface band from bulk bands while preserving all crystallographic spacetime symmetries of the surface,  such that the projector to the surface-band wave function is continuous  throughout $rBZ$, as illustrated in  \fig{fig:edge}(a-b).  (In contrast, the  surface states of a 3D Chern insulator\cite{kohmoto_3Dcherninsulator} are un-detachable.\cite{vanderbilt_book_berryelectronicstructure}) The conceptually simplest formulation of the surface Chern number is as a  $rBZ$-integral of the Berry curvature $\bcalf$ of the surface-band wave function: 
\e{ C_s:=\f1{2\pi}\int_{rBZ} Tr[\bcalf(\bk) \cdot \hat{\bn}] d^2k, \la{defineCs}}
with $\hat{\bn}$ the outward-normal vector for a chosen surface, and
$Tr$ a trace over all surface bands, \textit{independent} of electron filling. The second half of the Letter will present a more general formulation of $C_s$  -- via eigenstates of the projected position operator --  that is well-defined without need of detachment.



Let us demonstrate that $C_s$ is invariant under continuous deformation of the surface-terminated Hamiltonian, so long as both bulk gap and bulk translational symmetry are maintained. Since surface bands are detachable, their wave function is exponentially localized to the surface, and therefore their net Chern number cannot change under a bulk deformation \textit{alone}. What remains is to prove the absence of a continuous deformation between two Hamiltonians with the \textit{same} bulk but with distinct  surface Chern numbers: $C_s>C_s'$. Let us assume to the contrary that  such a deformation exists and is parametrized by $\lambda \in [0,1]$. It is simplest to view $C_s,C_s'$ as applying to the top surface of a finite slab, with periodic boundary conditions in two surface-parallel directions. As $\lambda$ is varied,  we hypothesize that $(C_s-C_s')$ quanta of Berry curvature can leave the top surface and enter the bulk.\cite{maryam_adiabaticpumpCSA,olsen_surfacetheoremCSA} Since the bulk Hamiltonians at $\lambda{=}0$ and $\lambda{=}1$ are indistinguishable, this Berry curvature must finally end up at the bottom surface of the slab. In the bulk perspective, there must then exist an adiabatic cycle for a bulk Hamiltonian  $H(k_1,k_2,k_3,\lambda)$ that is periodic in $\lambda$, such that $(C_s-C_s')$ quanta of Berry curvature is translated by a surface-normal lattice vector over one cycle. If this translation occurs through the low-energy (resp.\ high-energy)  band of $H(\bk,\lambda)$, then $(C_s-C_s')$ is identifiable as the second Chern number $\calc_2$ of the low-energy (resp.\ high-energy) band; $\calc_2$ is defined\cite{qi_topologicalfieldtheory} as the $(\bk,\lambda)$-integral of the second Chern character. The wrongness of our hypothesis is now self-evident, because $\calc_2$ necessarily vanishes for any band spanned by a single, analytic Bloch function of $\bk$.\footnote{The triviality of the first Chern class of a complex line bundle implies the triviality of the bundle.\cite{weil_varieties} See also Proposition 3.10 in \ocite{hatcher_vectorbundlesKtheory}}


The above argument relied on identifying $\calc_2$ as the number of Berry-curvature quanta pumped by one lattice vector in an adiabatic cycle. Indeed, if $H(\bk,\lambda)$ -- with trivial first Chern class and $\calc_2{\neq}0$ -- were diagonalized with open boundary conditions in the $z$ direction, then the boundary theory describes (3+1)D Weyl fermions with a net chiral charge\cite{nielsen_ninomiya} equal to $\calc_2$, in the three-dimensional, momentum-like space of ($k_1,k_2,\lambda$).\cite{olsen_surfacetheoremCSA} (These Weyl fermions may be derived in a continuum model -- as chiral zero modes of the (4+1)D Dirac equation with a domain wall.\cite{callanharvey_anomalies,qi_topologicalfieldtheory}) It is illuminating to view the boundary theory as (2+1)D fermions tuned by $\lambda$:  over one $\lambda$-cycle, $\calc_2$ quanta of Berry curvature must be transferred at critical closings of the boundary energy gap. Since the full Hamiltonian (that operates on both boundary and bulk Hilbert spaces) is periodic in $\lambda$, there must be a compensating pump of curvature through the bulk to exactly cancel the transfer of curvature at the boundary. Such a cancellation occurs between two systems of differing dimensionality, and has the same origin as the  Callan-Harvey anomaly inflow.\cite{callanharvey_anomalies,golterman_chiralfermions,stone_edgewaves} To our knowledge, the Callan-Harvey effect has previously been discussed  as an inflow of charge or energy-momentum, but never before as an inflow of Berry curvature. We have proven that $\calc_2{\neq} 0$ implies a Berry-curvature pump in the bulk; the converse statement -- that a pump implies $\calc_2{\neq}0$ -- has been proven by formulating the Chern-Simons magnetoelectric polarizability in terms of the Berry curvature of `hybrid Wannier sheets'.\cite{maryam_adiabaticpumpCSA,olsen_surfacetheoremCSA}  

Our argument for the invariance of $C_s$ shows that the Callan-Harvey effect is \textit{not} the origin of the Berry curvature on the Hopf-insulator surface. (In contrast, the Callan-Harvey effect is responsible for the nontrivial  surface anomalous Hall conductance of the `layered Haldane model'.\cite{olsen_surfacetheoremCSA})  Understanding the Hopf insulator not only reveals a new mechanism for Berry curvature to flow to the surface,  but also leads us to the advertised bulk-boundary correspondence $\chi{=}C_s.$ 

\begin{figure}
    \centering
    \includegraphics[width=\columnwidth]{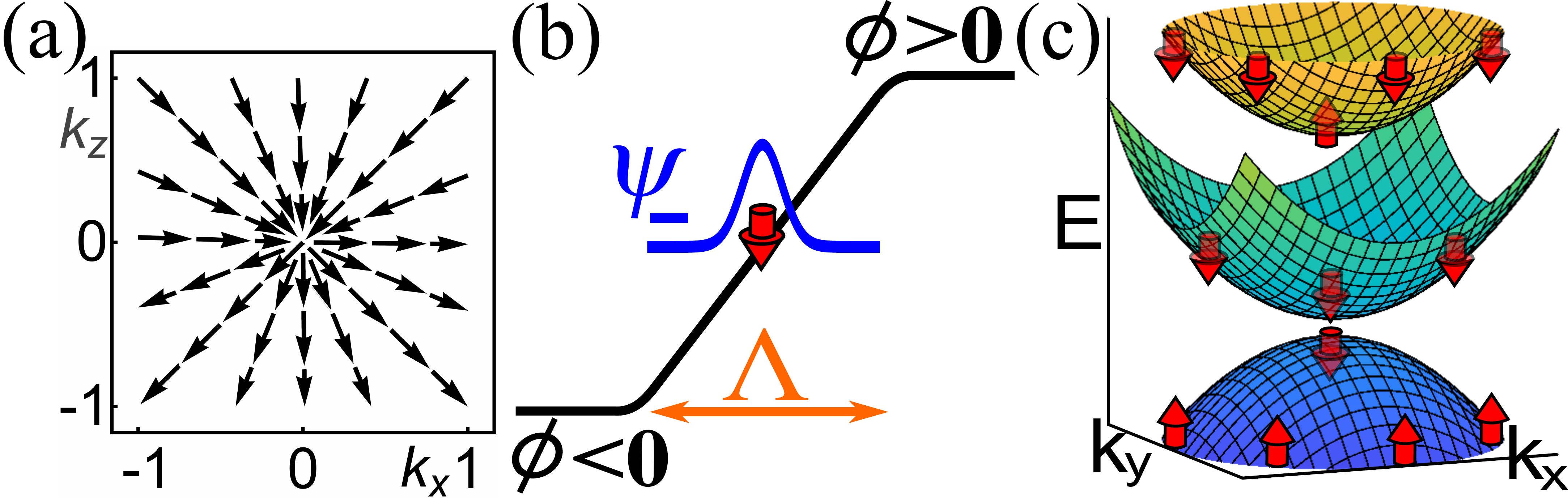}
    \caption{(a) Berry curvature ($F_1,F_3$) over a $k_2{=}0$ slice of the Berry dipole. (b) illustrates the spin-polarized wave function $\psi_{-}$ localized to a domain wall of $\phi(r_3)$. (c) contrasts the spin texture of the electron-like, domain-wall mode (middle band) with  those of the two bulk modes (top and bottom).  }
    \label{fig:domainwall}
\end{figure}

In the continuum perspective, the curvature may be understood from the perspective of non-chiral domain-wall modes of a generalized Weyl equation: $(i\partial_t -H)\psi=0$ with 
\e{ H(\bk) \eq -\bB\cdot \bsigma; \;\; \bB(\bk)=\dg{z}\bsigma z; \; \; \bsigma :=(\sx,\sy,\sz) \lin 
z(\bk)\eq (z_1,z_2)^T = (k_1 +i  k_2, k_3 +i \phi)^T. \la{dipolehamiltonian}}
We propose that this equation  describes the critical transition between a trivial and a Hopf insulator, in analogy with how the Dirac equation describes the critical point for an insulator with nontrivial Chern class.\cite{qi_topologicalfieldtheory} When $\phi{\in}\R$ is tuned to zero, both bands disperse quadratically in all momentum directions, and they touch at a point which acts as a dipolar source of Berry curvature, as illustrated in \fig{fig:domainwall}(a) by  a unit positive (resp.\ negative) Berry flux out of the southern (resp.\ northern) hemisphere.
Such a \textit{Berry dipole} contrasts with linearly-dispersing Weyl points that act as monopole sources.\cite{Nielsen_ABJanomaly_Weyl}
The {dipole} moment lies parallel to the $k_3$ axis, which is also an axis of rotation:
\begin{align}
{\pdg{U}_{\theta}}H(\bk){U^{-1}_{\theta}} = H(k_1',k_2',\pdg{k}_3), \;\; \pdg{U}_{\theta}=e^{i \theta \sz/2}    \label{fourfold}
\end{align}
with $k_1'=k_1\cos\theta-k_2\sin\theta$, and $U_{\theta}$ a spinor matrix representation with spin operator $\sz/2$. 
For $\phi{\neq}0$, $z/||z||\rightarrow \bB/||\bB||$ exhibits the standard Hopf map from $S^3\rightarrow S^2$, and the continuum analog of the Hopf invariant equals
\e{-\f1{4\pi^2}\int_{\R^3} \bF \cdot \bA \,d^3k= \f1{2}\text{sign}[\phi],\la{continuumcurvpol}} 
for \textit{both} the low-energy wave function ($z/||z||$) and its orthogonal complement ($\sy\bar{z}/||z||$).
$\phi{=}0$ is thus the critical point for a unit change in $\chi$ for both low- and high-energy bands. Away from criticality, the spin texture of the low-energy (resp.\ high-energy) band is skyrmionic (resp.\ anti-skyrmionic) over the $k_1k_2$-plane, as illustrated in \fig{fig:domainwall}(c). (In comparison, the meronic spin texture\cite{bernevig_hgte} of a $(2+1)D$ massive Dirac fermion is  `half' of a skyrmion.)

To manifest the nontrivial boundary mode that develops at criticality, we  solve  \q{dipolehamiltonian} with $k_3{\rightarrow}{-}i\partial_{r_3}$ and $\phi(r_3)$ having a domain-wall profile.  It is analytically convenient to choose  $\phi(r_3)=\phi'r_3$ that is linear in the interval $|r_3|{<}\Lambda$ (with $\Lambda {\gg} |\phi'|^{\minus 1/2}$), with $\phi$ tapering off to a constant outside this interval, as depicted in \fig{fig:domainwall}(b). Then a  Gaussian-localized, spin-polarized  mode $\psi_{\eta}$ exists with a Schr\"odinger-type dispersion $\var_{\eta}$: 
\e{ \psi_{\eta} =\kappa_{\eta}e^{-|\phi'|r_3^2/2},\; \;\var_{\eta}= -\eta(k_1^2+k_2^2-|\phi'|),\notag}
with $\sz{\kappa_{\pm}}{=}{\pm}{\kappa_{\pm}}$. The mass and spin depend on sign[$\phi']{:}{=}{-}\eta$; we refer to modes with positive (resp.\ negative) mass with respect to $(k_1,k_2)$ as electron-like (resp.\ hole-like). 



The contrasting spin textures -- constant for the domain-wall mode but (anti)skyrmionic for the bulk modes -- imply that the
integrated Berry curvature  vanishes for both electron- and hole-like domain-wall modes, but equals ${+}2\pi$ (resp.\ ${-}2\pi$)  for the electron-like (resp.\ hole-like) bulk mode. We will see that the \textit{difference} in curvature -- between electron-like bulk and electron-like domain-wall modes -- is maintained upon regularizing our continuum model on a lattice. The Hopf insulator is a lattice regularization  satisfying that all bulk modes have \textit{trivial} first Chern number; this would imply that the regularized domain-wall mode has a \textit{nontrivial} first Chern number  (${=}\eta$) which depends on the orientation of the domain wall, like how our proposed surface Chern number  depends on the surface orientation $\hat{\bn}$ [cf.\ \q{defineCs}]. 

Since $\phi{=}0$ describes a Berry dipole and not a monopole, a regularization can be found without any fermion doubling,\cite{Nielsen_absenceofneutrinos} meaning that the energy gap of the lattice model closes at a single isolated wavevector (the BZ center) when $\phi{=}0$.  
A lattice regularization satisfying the above-stated properties is given by the Moore-Ran-Wen tight-binding model,\cite{Hopfinsulator_Moore} though they did not emphasize the dipole picture that is presented here. Their lattice Hamiltonian has the form of  \q{dipolehamiltonian} with $z$ generalized to a $BZ$-periodic function with a Wilson coupling: $ z_1{=}\sin k_1{+}i\sin k_2, z_2{=} \sin k_3 {+}i(\phi+{\sum}_{j=x}^z\cos k_j -3).$ The tight-binding Hilbert space may be viewed as comprising spin-half particles ($\sz{=}{\pm}1$) in each unit cell of a tetragonal lattice. 



Let us prove $\chi{=}C_s$ for the Moore-Ran-Wen model. Despite the specificity of the model,  the correspondence would extend to  the entire homotopy class of translation-invariant, two-by-two Hamiltonians. This is because $\chi$ and $C_s$ are individually invariant under any continuous, bulk-gap-preserving deformations that preserve the bulk-translational symmetry; this fact is well-known for $\chi$\cite{pontrjagin_classification,Hopfinsulator_Moore} and has been proven above for $C_s$. 
Hamiltonians with nonzero $\chi$ may have any non-translational symmetry that does \textit{not} invert the pseudoscalar: $\chi {\rightarrow}{-}\chi$. Particularly, the Moore-Ran-Wen regularization retains an order-four rotational symmetry [cf.\ \q{fourfold}], which not only allows us to utilize our rotation-symmetric domain-wall analysis, but also manifests several features unique to the \textit{crystalline Hopf insulator}.



\begin{figure}
    \centering
    \includegraphics[width=\columnwidth]{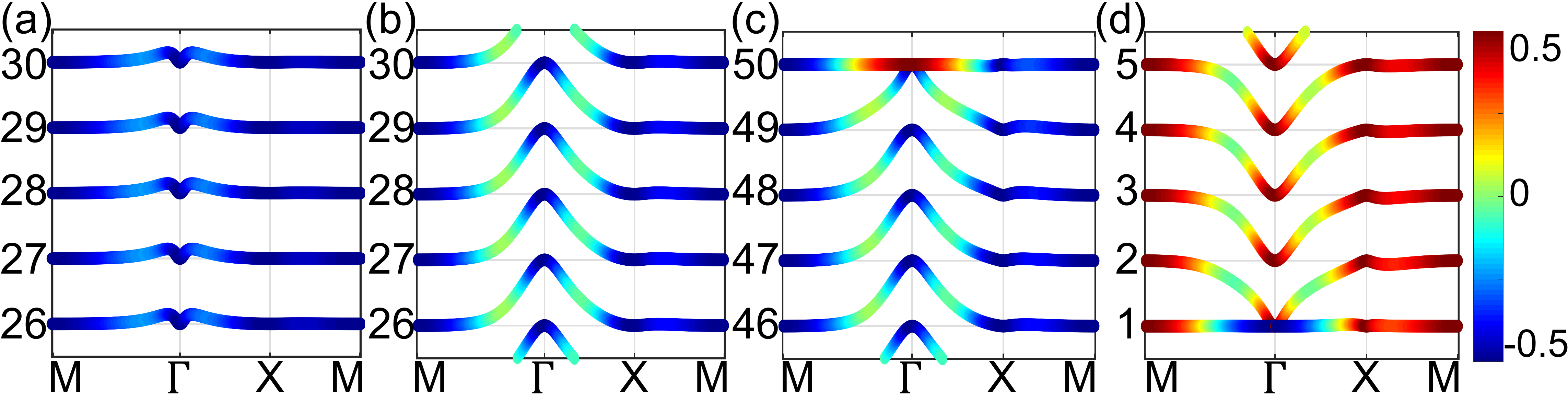}
    \caption{Dispersion of the projected-$z$ operator  over a high-symmetry line in $rBZ$, for a $50$-cell-wide 001 slab. $z$  is projected to the low-energy band of the Moore-Ran-Wen model in (a-c), and to the high-energy band in (d). $\phi{=}{-}0.2$ in (a), and ${=}1$ in (b-d). (a-b) illustrate the bulk dispersion, while (c-d) depict the top and bottom surfaces respectively. The expected spin $\langle \sz/2\rangle$ is depicted according to a color scheme on the right. }
    \label{fig:wanniersheets}
\end{figure}

For sufficiently negative $\phi$, $H(\bk){\approx}\phi^2\sz$ implies that both low- and high-energy bands are spin-polarized and the Hopf invariant vanishes.
It is instructive to represent both  bands by basis functions that are extended as Bloch functions with wavevector $(k_1,k_2){\in}rBZ$, but exponentially-localized in the $z$-direction as Wannier functions.\cite{AA_wilsonloopinversion,alexey_wannierrepZ2TI,Maryam_wanniercentersheets} In units of the lattice period, the positional center $\bar{Z}$ of a hybrid function  is related\cite{zak_berryphase} to the Berry phase as
$\bar{Z}(k_1,k_2) {=}  \int_{0}^{2\pi} A_z(\bk)dk_3/{2\pi}.$
Owing to the discrete translational symmetry along $z$, the \textit{hybrid centers} $\bar{Z}$  are arranged as a Wannier-Stark ladder\cite{wannier_starkladder,nenciu_review,TBO_JHAA} with unit spacing between adjacent rungs. The  low-energy band having only spin-down character implies the Berry connection can be trivialized, implying $\bar{Z}=0$ (modulo integer) for all $(k_1,k_2)$. For finite $\phi{<}0$, the low-energy band acquires some spin-up character \textit{except} at the four rotation-invariant $(k_1,k_2)$-points [denoted as ${\Gamma},{X},{Y},{M}$ in \fig{fig:wanniersheets}(a)], where distinct rotational representations cannot mix. Despite the nonzero dispersion, each hybrid function retains the topology of a `flat sheet' which is pinned to integer values at the rotation-invariant $(k_1,k_2)$-points.

As negative $\phi$ is tuned through zero, the spins of the bulk eigenstates of $H(\bze)=\phi^2\sz$ are  \textit{not} inverted; nevertheless, the Hopf invariant increases by unity [cf.\ \q{definechi}]. Analogously, the Berry phase of the low-energy mode in the continuum model is $\int_{\R} A_z(0,0,k_3) dk_3=\pi \,\text{sign}[\phi]$, which implies a $2\pi$-quantum of Berry phase is transferred from the high- to low-energy band (at the $rBZ$-center). Therefore, there must  be a discontinuous change in the connectivity of the Wannier-Stark ladder, resulting in upward-protruding (resp.\ downward-protruding) sheets for the high-energy (resp.\ low-energy) band, as illustrated in \fig{fig:wanniersheets}(b-d).  The triviality of the first Chern class ensures the periodicity of $\bar{Z}$ over $rBZ$, hence no net charge flows if $(k_1,k_2)$ is varied along ${\Gamma}M{\Gamma}$. Despite the absence of a conventional Thouless pump over an adiabatic cycle,\cite{thouless_pump} observe that the hybrid center  is translated by exactly one lattice period at the half-way mark before returning to its original position. Such a \textit{returning Thouless pump} implies that the Wannier function -- obtained by one-dimensional Fourier transform of the hybrid Bloch-Wannier functions -- cannot be localized to a \textit{single} lattice site. Such a topological obstruction is a necessary condition for a quantized, \textit{real-space} polarization of Berry curvature. (There is, however, no obstruction to \textit{exponentially}-localized Wannier functions respecting all crystallographic spacetime symmetries, according to a theorem proven by one of us in \ocite{nogo_AAJH}.)

The above-described connectivity of Wannier-Stark sheets holds for an infinite ladder, but is frustrated for a finite slab with  rotation-invariant surface terminations -- meaning that protrusions of the lowermost (resp.\ uppermost) sheet of the high-energy (resp.\ low-energy) band are incompatible with the confining surface potentials. These two frustrated sheets are continuously deformable to the surface bands of the finite-slab Hopf insulator.\cite{fidkowski_bulkboundary,Cohomological} For these surface bands, the spin at the $rBZ$-center is undetermined from a bulk analysis but determined by our domain-wall analysis, which asserts that the electron-like (resp.\ hole-like) mode localized to the bottom (resp.\ top) surface has spin down (resp.\ up); this is confirmed by a numerical calculation illustrated in \fig{fig:wanniersheets}(c-d). A colloquial interpretation is that the lowermost sheet of the high-energy band `acquires' a spin-down particle from the low-energy band, which has been `left over' at the bottom surface due to a returning Thouless pump in the upward direction. Effectively, there is a mutual exchange of rotational representations between surface bands localized to opposite surfaces of a finite slab, manifesting a type of symmetry `teleportation'. 


The unit gain of angular momentum for the top-surface band introduces an irremovable twisting of the wave function that is quantified by the Chern number $C_s=1$.\footnote{The exact value for $C_s$ is obtained by applying an identity in \ocite{chen_TSCprojective}} Bearing in mind that the surface-normal vector $\hat{\bn}$ flips between top and bottom surface [cf.\ \q{defineCs}], we arrive at  $C_s=\chi=1$ for \textit{both}  surfaces.  It is straightforward to generalize this equality for any integer value -- by substituting $k_3 \rightarrow N k_3$  ($N\in \Z$) in the Moore-Ran-Wen model, such that   $\phi=0$ marks a critical point  with  $|N|$  Berry dipoles. The nontrivial Berry curvature implies that the full subspace of surface states (on a single surface) cannot be realized by any strictly-2D, tight-binding lattice model. This is analogous to how chiral fermions on $\Z_2$-topological-insulating boundaries\cite{callanharvey_anomalies} evade the fermion-doubling theorem.\cite{nielsen_ninomiya,nielsen_ninomiya_II}

Our Chern-number analysis may be verified by detaching surface bands and numerically calculating $C_s$ through \q{defineCs}. It is possible to formulate $C_s$ without detaching surface bands, but requiring the milder condition that a spectral gap exists throughout $rBZ$ and separates two orthogonal subspaces $P$ and $Q$. Such a gap always exists for certain surface terminations of the considered class of Hamiltonians,  because the surface bands are continuously deformable to hybrid Bloch-Wannier bands which disperses like `flat sheets'.\cite{fidk2011,Cohomological} (This `sheet' topology applies to all surfaces of the Hopf insulator, implying that a finite-size, Hopf-insulator crystal can in principle be truly insulating -- in both bulk and surface.) Let us consider a semi-infinite geometry that is periodic in $x$ and $y$, and with the position operator $z$ taking only negative values. We then diagonalize the projected position operator $PzP$, and define $C_s^P[n]$ as the first Chern number of the $n$ Bloch-Wannier eigenbands  whose eigenvalues lie closest to $z{=}0$. Viewed as a sequence in $n$, $C_s^P[n]$ has a unique accumulation point ($:=C_s^P$) because all bulk eigenbands have trivial first Chern class. Analogously defining $C_s^Q$ as the accumulation point for the orthogonal subspace $Q$, the surface Chern number is then $C_s{:=}C_s^P{+}C_s^Q$. This formulation emphasizes that $C_s$ is uniquely defined, e.g., it is manifestly invariant if  a Chern-insulating layer is adsorbed onto the surface. Practically, one may calculate $C_s^P{+}C_s^Q$ by summing Chern numbers (of Bloch-Wannier eigenbands) on \textit{half} of a sufficiently wide slab.




What does $\chi{=}C_s$ imply for the magnetoelectric response of the Hopf insulator? It is known that the geometric contribution to the  frozen-lattice, orbital magnetoelectric polarizability equals $(\theta/2\pi$ mod $1)e^2/h$, with the `axion angle'\cite{Wilczek_axion} $\theta$  equal to an integrated Chern-Simons three-form of the Berry connection.\cite{qi_topologicalfieldtheory,essin_magnetoelectric} For the considered class of Hamiltonians, $\theta$ is gauge-invariant and simplifies to $\pi\chi$, thus the Hopf insulator with odd $\chi$ exemplifies the simplest axion insulator\cite{qi_topologicalfieldtheory,essin_magnetoelectric,turner_inversionsymmetricTI,hughes_inversionsymmetricTI,vanderbilt_book_berryelectronicstructure,Schindler_higherorderTI}  with only a Pauli-matrix Hamiltonian. The possibilities for surface anomalous Hall conductance  on different facets of a finite-size, Hopf-insulator crystal are rich, and depend on the electrochemical potential as well as the energy dispersion of surface bands. 
Generically surface bands are partially filled, such that each facet behaves like a 2D anomalous Hall metal.\cite{haldane_berryonfermi} The differing crystallographic symmetries on distinct facets allow for non-identical fillings, e.g., it is possible for the rotation-invariant 001-surface bands of the Moore-Ran-Wen model to be completely filled, with the rotation-variant side-surface bands completely depleted -- this leads to $|C_s|$ number of chiral modes\cite{jackiwrebbi_solitons,goldstonewilczek_solitons,fanzhang_surfacestatemagnetization,sitte_TIinBfields} that run around the edges of the 001 facet. Such a scenario is realizable if the electrochemical potential on distinct surfaces are tunable, perhaps by differential doping or gating. 

Our last comment is aimed toward widening the search for solid-state realizations of the Hopf insulator, which would complement existing ultracold-atomic platforms.\cite{schuster_hopfdipolarspin,Deng_probeknotshopfins}  The first solid-state proposal of the Hopf insulator  -- on a distorted pyrochlore lattice with non-collinear magnetic order and strong spin-orbit coupling -- has not been realized in twelve years.\cite{Hopfinsulator_Moore} It is not recognized by the original authors that the Moore-Ran-Wen tight-binding model is also compatible with an integer-spin representation of the black-white magnetic space group P4m'm', which includes not only four-fold rotation but also magnetic reflections. This symmetry representation applies to a parity-breaking ferromagnet with {negligible} spin-orbit coupling; the two relevant orbitals (per unit cell) differ in angular momentum by $\hbar$.\\

\begin{acknowledgments}
\noindent \textit{Acknowledgments} We thank Joel Moore, Eduardo Fradkin and Michael Stone for their expert advice on the Hopf insulator, anomalies and characteristic classes, and Zhu Penghao for informative discussions on the non-geometric magnetoelectric polarizability. We would also like to thank Barry Bradlyn, Ken Shiozaki, Judith H\"oller, Nicholas Read, Ricardo Kennedy, G.~M.~Graf, I.~Souza, T.~Pahomi, D.~Gresch for useful discussions. AN and AAS acknowledge the support of the SNF Professorship grant along with SNSF NCCR MARVEL and QSIT programs.  AA was supported initially by the Yale Postdoctoral Prize Fellowship, and subsequently by the Gordon and Betty Moore Foundation EPiQS Initiative through Grant No. GBMF4305 at the University of Illinois. 
\end{acknowledgments}
 
\bibliography{bib_Apr2018}

\end{document}